# ADAPTIVE LEARNING EXPERT SYSTEM FOR DIAGNOSIS AND MANAGEMENT OF VIRAL HEPATITIS

Henok Yared Agizew

Department of Management Information System, Mettu University, Mettu, Ethiopia.

**ABSTRACT**

*Viral hepatitis is the regularly found health problem throughout the world among other easily transmitted diseases, such as tuberculosis, human immune virus, malaria and so on. Among all hepatitis viruses, the uppermost numbers of deaths are result from the long-lasting hepatitis C infection or long-lasting hepatitis B. In order to develop this system, the knowledge is acquired using both structured and semi-structured interviews from internists of St.Paul Hospital. Once the knowledge is acquired, it is modeled and represented using rule based reasoning techniques. Both forward and backward chaining is used to infer the rules and provide appropriate advices in the developed expert system. For the purpose of developing the prototype expert system SWI-prolog editor also used. The proposed system has the ability to adapt with dynamic knowledge by generalizing rules and discover new rules through learning the newly arrived knowledge from domain experts adaptively without any help from the knowledge engineer.*

## 1. INTRODUCTION

Viral hepatitis is an infection that results for liver damage. This inflammation of liver is caused by five different viruses such as hepatitis viruse A, B, C, D and E. Those five different types of viruses have the ability to cause acute disease, but the maximum number of deaths throughout world result from chronic hepatitis virus B or hepatitis virus C infection (WHO, 2013).

According to the Annual Report of WHO (2017), Hepatitis B is highly endemic and probably affects an estimated 5–8% of the population in Africa, mainly in West and Central Africa. It is estimated that around 19 million adults are infected with chronic hepatitis C in this region. The growing mortality rate of peoples living with HIV is caused by viral hepatitis. As a result, nearby 2.3 million people living with HIV are also infected with hepatitis virus C and about 2.6 million people infected with hepatitis virus B.

Expert System is a system that contain tacit and explicit knowledge (Wielinga et al., 1997). It is a system that can provide valuable knowledge for the sake of solving different problems in defined application domain (Kasabov et al., 1996). It has been practical successfully in almost every field of human activities due to it's ability to represent, accommodate and learn knowledge; it is also capable of taking decisions and communicating with their users in a responsive or friendly way (Wielinga et al., 1997). Through using such system it is possible to get so many benefits like wise decisions, learning from experience, explanation and reasoning, and solving problems (Kock, 2003).

Intelligent expert systems can be used to exploit productivity, capturing scare knowledge to aid document rare knowledge and enhance capabilities of problem solving in the most crucial way to eliminate the mortality rate and wastage of time to see the physician (Wan-Ishak et al., 2011). This system developed by imitating human intelligence could be employed to help the medical doctors in making decision without consulting the specialists. The system is not meant to replace the specialist or medical doctor, but it is developed to help general physicians and nurses in diagnosing, predicting condition of patient, and providing treatments based on certain grounds or experiences.

In order to develop such medical expert systems, adaptive learning mechanisms used the input sign or symptoms of the disease, the current available knowledge base, and the valid diagnosis result to produce an updated knowledge base. Before applying the adaptive learning techniques for the expert system, deciding how and when to apply those different types of learning mechanisms is very important. These mechanisms may refine rule statistics, rule hypotheses, or add new rules.

## 2. STATEMENT OF THE PROBLEM

Chronic viral hepatitis is now the second biggest killer after tuberculosis. About 325 million people are caused by chronic viral Hepatitis B and Hepatitis C throughout the world. This is 10 times greater than the global epidemic of HIV. More than 3600 people are die every day in case of the viral hepatitis disease (WHO, 2017).

According to Dhiman (2018), more than 70 million Africans are affected by chronic diseases like viral hepatitis B and C. Which means, about 60 million people are affected with hepatitis B and about 10 million people affected with hepatitis C. The disease affects adults and the productive young people's most of the time. As a result, the financial problem could be raised highly in this region. Moreover, Ethiopia is regarded as a country with out any controlling and management strategy for the viral hepatitis. However, the country is categorized under the regions with intermediate to hyperendemic viral hepatitis infections (WHO, 2013).

Developing expert systems would play essential and critical role in providing aid for giving diagnosis, capturing guidelines and knowledge and experience of well educated and experienced physician and make it available for those health workers which also help the country in providing a quality health care service for the people who are living in remote and rural areas. Several studies were conducted on knowledge based systems in order to support reasoning and finding solutions for certain problems like in the area of health and medicine. However, the knowledge based system implemented using rule-based representation technique did not hold all the necessary rules i.e., they do not have an ability to discover new rules and generalize the existing rules in order to adapt the new environment dynamically.

Therefore, this study aims to develop adaptive learning expert system that can improve the compresivenes of advice provided by the system in line with accurate diagnosis and management of viral hepatitis using both discovery and generalization mechanis.

## 3. RESEARCH QUESTION

In this regard, this study explores and finds solutions for the following research questions:

> ➢ What suitable domain knowledge available for the diagnosis and management of viral hepatitis ?

> ➢ What are the suitable ways to develop Adaptive learning expert system for diagnosis and management of viral hepatitis ?

## 4. OBJECTIVES

The general objective of this study is to develop adaptive learning expert system that can provide advice for physicians in order to aid the diagnosis and management of viral hepatitis. Specifically:

- ➢ To acquire the required knowledge from the human expert and documents
- ➢ Modeling and representing the knowledge in a suitable way for knowledge based expert system implementation
- ➢ Designing the prototype expert system with adaptive learning ability

## 5. SCOPE AND LIMITATION

In order to developing expert systems, several approaches are available. However, for the sake of implementing this adaptive learning expert system for diagnosis and management of viral hepatitis the rule based reasoning approach is used. The system takes new cases of the problem and refining existing rules body, generalizing rules and descovering new rules to provides accurate result and support for decision making. There are five (5) different types of viral hepatitis but including all viral hepatitis types require further investigation and influence the system performance. Due to this reason this study focused on chronic viral hepatitis (viral hepatitis B and viral hepatitis C) only.

## 6. LITERATURE REVIEW

The availability of advanced computing facilities and other resources, currently the attention of the knowledge engineer is changed to the more demanding activities, which might require proper intelligence. Knowledge based system (KBS) can act as an expert on demand without wasting time and being anywhere, therefore the society could solve the decision making problem easily by using such expert system (Sajja & Akerkar, 2010).

Both Expert system (ES) and knowledge based systems (KBS) are terms that are used interchangeably in machine learning technology (Heijst, 2006). KBS is one of the major family members of the AI group and Expert systems provide ability to act as domain expert. expert system also used to store the knowledge of different domain experts for long time (Malhotra, 2015).

### 6.1 THE PROCESS OF DEVELOPING EXPERT SYSTEM

The process of knowledge engineering follows its own basic procedures. The main phases of the expert system development processes are planning, knowledge acquisition, knowledge representation and evaluation (Sajja & Akerkar, 2010).

The knowledge of the expert is stored in his mind in a very abstract way. Also every experts may not be familiar with expert systems terminology and the way to develop an intelligent system. Knowledge Engineer is the responsible person to acquire, transfer and represent the experts' knowledge in the form of computer system. But, the remains are users' of the knowledge based systems (Sajja & Akerkar, 2010).

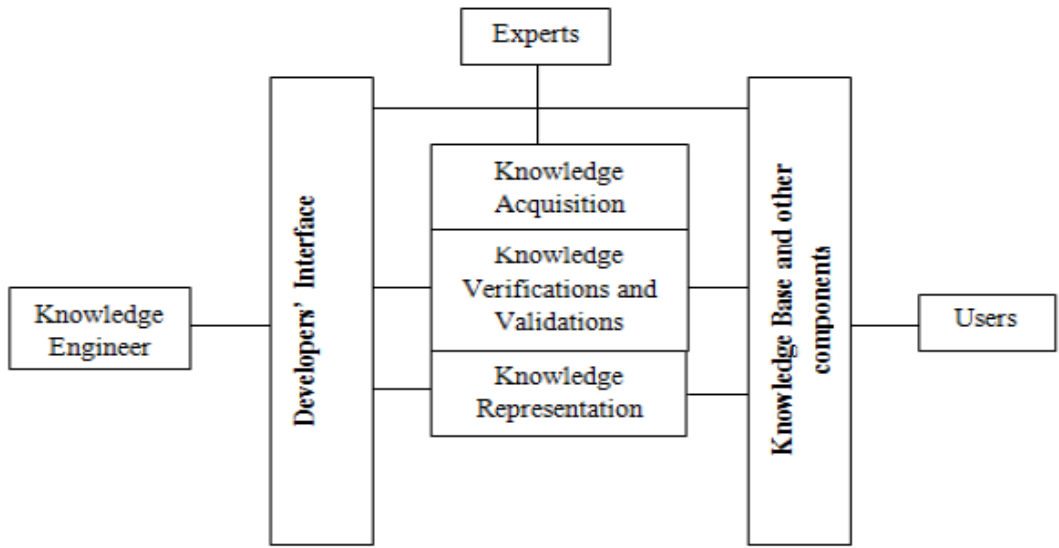

Figure 1.6.1 Development of expert System (Sajja and Akerkar, 2010).

## 6.2 EXPERT SYSTEM REASONING TECHNIQUES

An expert system is a computer program that represents and reasons with knowledge of some specialist subject with a view to solving problems or giving advice. To overcome problems occurs in the domain expert level, expert systems will require valuable access to a substantial domain knowledge base, and a suitable reasoning techniques to apply the knowledge in to the problem they are given (Jackson, 1999). The most widely used expert system reasoning methods are either rule-based or case based or a hybrid of them (Leake, 1996). Among those different reasoning methods, the rule-based reasoning approach is used for the purpose of conducting this reseasrch.

## 6.3 RULE-BASED REASONING

Rule based reasoning is a system whose knowledge representation in a set of rules and facts. One of the most popular knowledge representation and reasoning techniques are using symbol as a rule to represent propositions and relations in order to assist reasoning. This popularity is mainly due their naturalness, which facilitates comprehension of the represented knowledge. A rule can be supposed of as a condition action couple. Any rule has two different sections such as; the IF section which is also known as antecedent or premises or condition and the remaining section is the THEN section which also known as consequent or action or conclusion. Those yields the format shown below;

                **IF** condition holds
                **THEN** do action

The conditions are fired or not rendering to what fact is existing currently. If available situations or facts of a rule are fulfilled, then the action is created and the rule is said to be executed. The way of representing the real world situations like troubleshooting, advisory, diagnosis and treatment as a rule is easy rather than others. As a result, so many works were done by using the rule based reasoning techniques to represent the existing domain knowledge.

### 6.4 EXPERT SYSTEM DEVELOPMENT APPROACH

The expert system development methods can be classified into three categories: manual, semi-automatic, and automatic (Gamble,et al.,2003).

- Manual Methods: in this manner, the knowledge is gathered from the domain experts and also extracted from other knowledge sources through interview by the knowledge engineer to codes it in the knowledge base in order to develop the expert system. There are different types of the manual knowledge gathering methods, however the three main manual methods are interviewing the domain experts, tracking the process of reasoning, and observation (Gamble, et al.,2003). Manual methods has the drawbacks such as slow processing speed, very expensive to collect appropriate knowledge and the accuracy of gathered knowledge. Therefore, there is a trend toward automating the process as much as possible. For instance, KBS which requires everything from experts and knowledge engineers are an example of manual methods.

**Semi-Automatic Methods: are divided into two categories.**

1. The first one is; the domain experts could build the knowledge bases without any support or little help of the knowledge engineer.
2. Second; intended to support knowledge engineers in order to execute the necessary tasks efficiently and effectively. For instance, KBS which are learning new knowledge from experts and did not require knowledge engineers to update or maintain the system are an example of semi-automatic methods.

- Automatic Methods: in this section, it is said to be the system need a little help or does not require any contribution from domain experts and also from the knowledge engineers, because the system update it selves automatically. For example, the induction method, which generates rules from a set of known cases, can be applied to build a knowledge base. The role of the expert and knowledge engineers are minimal. The term automatic may be misleading, but it indicates that, compared with other methods, the contributions from a knowledge engineer and an expert is relatively small.

For instance, the expert system with learning ability is an example of automatic methods because they are able to add new knowledge and maintain the system with minimum contributions from a knowledge engineer and an expert.

### 6.5 EXPERT SYSTEM DEVELOPMENT TOOLS

In the process of expert system development the software and hardware tools are very crucial in as well as both general purpose programming languages like java and framework like .NET and useful specific purpose ready made utilities such as Expert System Shell (JESS), GURU, and Vidwan. In most cases tailor made expert system can be developed using programming languages like LISP (List Processing) and Prolog (Nilsson & Maluszynski, 2000). For the purpose of conducting this research work the Logic Programming language with SWI Prolog editor tool is used.

**SWI Prolog:** SWI Prolog is the most comprehensive, user-friendly and widely used Prolog development environment. It has good development facilities and aids a graphical debugging environment to apply the graphical user-interface feature easily in the implementation of the expert system (Wielemaker, 2010). Moreover, SWI Prolog is well maintained and available for all major platforms such as Windows, Mac OS, Linux, etc (Wielemaker, 2010).

## 6.6 RELATED WORKS

Santosh et al.,(2010) developed "an expert system for diagnosis of human diseases". The system is rule-based system and makes inferences with symbols for knowledge representation. They recommended automated diagnosis system should give explanation for the conclusion, a factor which is important for user acceptance for the sake of testing the accuracy of the diagnosis provided by developed expert system.

The study conducted by Seblewongel, (2011) "A prototype of rule-based KBS for anxiety mental disorder diagnosis". This study was conducted using rule-based representation technique to find solution for mental health problem which is not gaining enough cares. Thus, knowledge about anxiety mental disorder was acquired using both structured and unstructured interviews from domain experts and from secondary sources by using document analysis method to find solution of the problem. For modeling the knowledge gathered before, the decision-tree modeling technique is used to properly structure the rules and procedures existed in the domain area. Finally, the system is developed using manual KBS development method with SWI Prolog editor tool.

The study conducted by Solomon,(2013),"Aself-learning knowledge-based system for diagnosis and treatment of diabetes". The study focused on memorizing or remembering patients past history. It is rule based system with backward chaining technique and the system is developed using semi-automatic KBS development method with SWI Prolog editor tool.

Paulo and Augusto, (2018), tried to develop the mobile system that can support prediction of childrens with autism. To developing the system the idea behind fuzzy neural networks were used with the concept of AI and machine learning. The system is capable to provide some sort of question and immidiate respones for users through using binary classification labels.

## 7. KNOWLEDGE ACQUISITION, MODELLING AND REPRESENTATION

According to Sajja et al.,2009, Knowledge Acquisition is the way of gathering domain area knowledge using different techniques such as interview, observation and questioner for gathering domain experts knowledge and using document analysis technique for gathering knowledge from other sources of information such as books, databases, guidelines, manuals, journal articles and computer files. Knowledge acquisition is the first step and critical task in the development of knowledge based system.

In this study, to acquire the needed knowledge, both secondary and primary sources of knowledge are used. The primary data (tacit knowledge) gathered from experts in the domain area (Internists of the St.Paul Hospital, Ethiopia) using structured and semi-structured interview. Secondary sources of knowledge have been gathered from the Internet sources, viral hepatitis diagnosis guide line, viral hepatitis prevention, viral hepatitis treatment guide line, manuals, and published Articles.

### 7.1 CONCEPTUAL MODELING

Modeling should be the perquisite activity for implementing expert systems. Models are applied to acquire the important characteristics of problem domains by decomposing them into more controllable parts that are simple to know and to use. In this study, decision tree is used to model the levels of decisions that domain experts use during diagnosis and management of viral hepatitis. The details of the conceptual model using decision tree are depicted in figure 6.2.1 and figure 6.2.2 below.

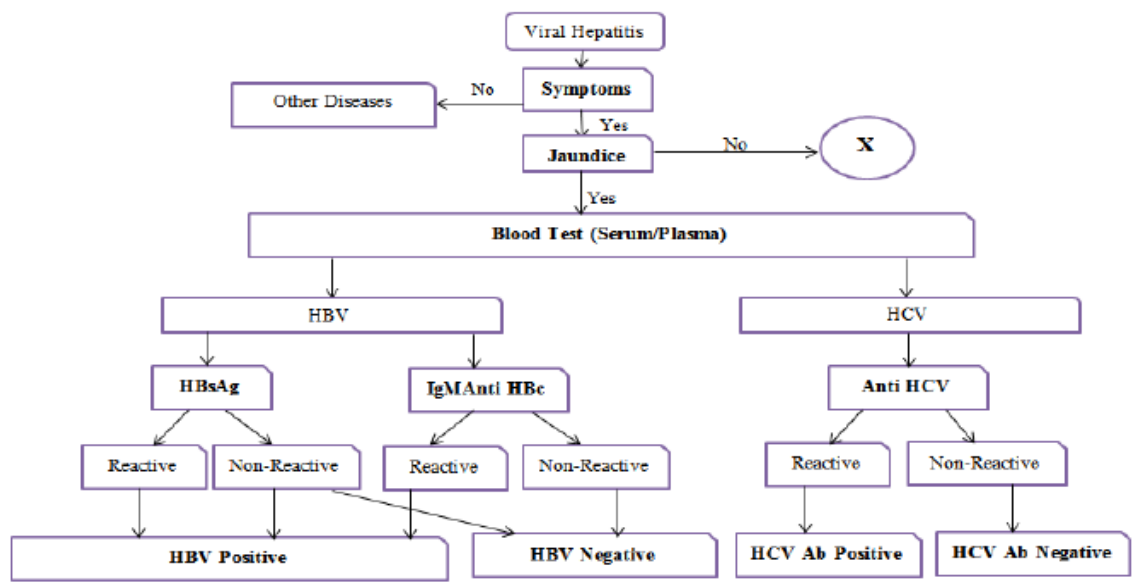

Fig 6.2.1 : Decision tree for diagnosis of HBV and HCV (with Jaundice)

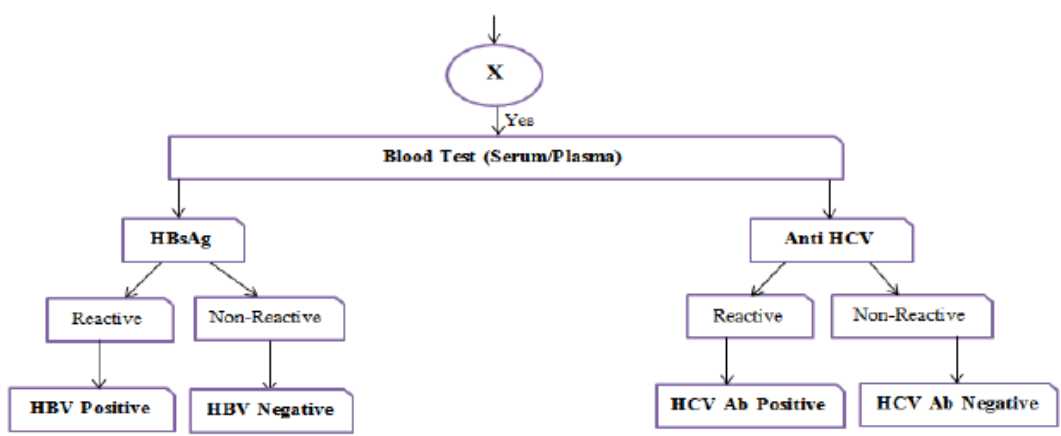

Fig 6.2.2: Decision tree for diagnosis of HBV and HCV (with out Jaundice)

**7.2 KNOWLEDGE REPRESENTATION**

Once the model of the proposed system is designed, the next stape will be representing the knowledge to make the coding phase easy for the Knowledge Engineer. For the purpose of conducting this research work, the rule based reasoning technique with both forward and backward chaining is used.

In the rule-based reasoning the domain knowledge that acquired from the domain experts and related medical documents are represented as a set of premises and condition which consisits both facts and rules in the knowledge base. In order to develop the fully functional and expert system with the accurate advices and adaptive learning ability the rule could plays a crucial role. For instance; the HBV and HCV diagnosis knowldege in the decision tree above is represented as the following manner;

**Knowledge Representation for HBV Diagnosis with Jaundice**

*If*
*the patient has viral hepatitis sympthoms,*
*with jaundice,*
*Then*
*Check the HBV blood tests with HBsAg and IgM Anti HBc.*
*If HBsAg result is Reactive,*
*Then HBV result is Positive.*
*If HBsAg result Non-Reactive,*
*Then HBV result is Positive.*
*If HBsAg result is Reactive AND IgM Anti HBc result is Reactive, Then HBV result is Positive.*
*If HBsAg result is Reactive AND IgM Anti HBc result is Non-Reactive, Then HBV result is Positive.*
*If HBsAg result is Non-Reactive AND IgM Anti HBc result is Non-Reactive,*
*Then HBV result is Negative.*
*Knowledge Representation for HCV Diagnosis with out jaundice*
*If*
*the patient has viral hepatitis sympthoms,*
*with out jaundice,*
*Then*
*Check the HBV blood tests with antiHCV.*
*If anti HCV result is Reactive,*
*Then HCV result is Positive.*
*If anti HCV result is Non-Reactive,*
*Then HCV result is Negative.*

## 8. ADAPTIVE LEARNING MECHANISM IN RULE-BASED EXPERT SYSTEM

According to Clair et al., (1987), Adaptive learning strategies classified into six basic categories. Those are strengthening, weakening, unlearning, generalization, discrimination, and discovery. Among those different types of adaptive learning strategies; generalization and discovery mechanisms are used in this research.

### 8.1 GENERALIZATION MECHANISM

Generalization mechanism has a crucial roles in the process of applying adaptive-learning techniques to the expert system. The main concepts behind this mechanism are minimizing the number of premises and deriving less restrictive hypotheses (Clair, 1987). Accordingly, the advantages of using this mechanism are (Clair, 1987):

- The derived rule could be applicable in most of the available environments
- Identifying the difficult condition in which a rule should have fired didn't execute easily.

Here, is the rule representation for the rule generalization mechanism in adaptive learning mechanism.

**Rule 1:**

If
the patient has viral hepatitis sympthoms,
with jaundice,
HBsAg result is Reactive,
HBsAg result Non-Reactive,
IgM Anti HBc result is Reactive,
Then
HBV result is Positive.

**Rule 2:**

If
the patient has viral hepatitis sympthoms,
with jaundice,
HBsAg result is Reactive,
Then
HBV result is Positive.

**Rule 3:**

If
the patient has viral hepatitis sympthoms,
with jaundice,
HBsAg result is Non-Reactive,
Then
HBV result is Positive.

**Rule 4:**

If
the patient has viral hepatitis sympthoms,
with jaundice,
IgM Anti HBc result is Reactive,
Then
HBV result is Positive.

In this example; Rule 1 is the generalized rule of Rule 2, Rule 3 and Rule 4. All three rules will fire whenever Rule1fires, so that the generalized Rule 1 should be used to replace the remaining rules.

The figure shown below is the system knowledge base where the knowledge gathered from different hepatitis sources are stored on it. This knowldge base is constructed from 33 (thirty three) different clouses and consists the diagnosis knowldge of viral hepatitis which are used to support physicians to identify weither the diagnosed case is positive or not. However, the following basic produre should be followed, in order to know which rule among the following rules has the maximum number of execution experience.

International Journal of Artificial Intelligence and Applications (IJAIA), Vol.10, No.2, March 2019

Figure 8.0. Viral Hepatitis Knowledge Base for indicating rule experiences

1. Reconsult the knowledge base of viral hepatitis at run time
2. Induce the rule to indicate each rule's execution experiences
3. Use the rule with the maximum number of execution exeperience to generalize the rule

Accordingly; the rule induction algorithm were used to proced the process of indicating the rule's execution experiences. Finally, results generated from the above sample cases of hepatitis knowledge following those three steps are shown below;

Figure 8.0.1 Rule execution experience indicator result

As a result, here is shown the condition and action pair of sample hepatitis diagnosis knowledge indiuced from the knowledge base above. For instance;

**Rule1:** hbscagreact = yes

igmantihbcreact = yes => [negative/1]
i.e. IF HBsAg result is Reactive AND IgM Anti HBc result is Reactive THEN the HBV result is Negative. This [negative/1] indicates as the rule 1 is fired or executed only one time. While,

**Rule2:** hbscagreact = yes
igmantihbcreact = no => [positive/9] , this also represents;

IF HBsAg result is Reactive AND IgM Anti HBc result is Non-Reactive THEN the HBV result is Positive. This [positive/9] also indicating as the rule2 is fired or executed nine (9) times. Accordingly, among those two different rules the rule that has more number of execution time is rule2 becouse it is fired 9 times while rule1 is fired once. Therefore, the expert system used rule2 which can generalize the rule1 in order to provide accurate diagnosis result for physicians. Those remaining other rules are also processed in the same manner.

**8.2 DISCOVERY MECHANISM**

Discovery mechanism is the process of discover new rules through utilize input symptoms and the current knowledge base in order to adapt with the new situation. This discovery mechanism is very important whenever the incomplete result is provided by the system and when the system does not have a response to the new occur situations. The figure below shows how to discover the new rule in prolog programming language.

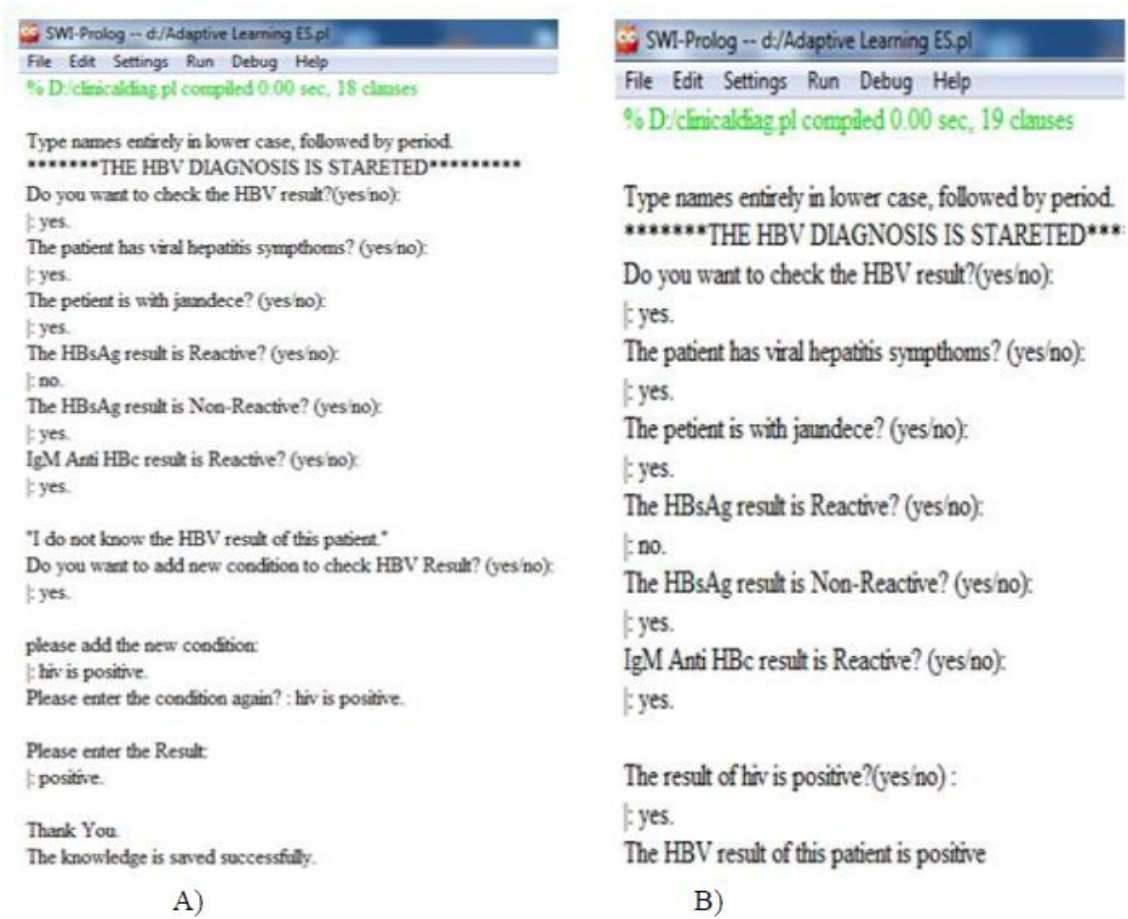

Figure 8.1: Discovery mechanism sample code.

International Journal of Artificial Intelligence and Applications (IJAIA), Vol.10, No.2, March 2019

The knowledge gathered from the domain experts is coded in to the knowledge base of the developed expert system as shown above. Also the validated knowledge base automatically replaces the previous value existed in the knowledge base in order to adapt with the new situation.

The figure 1.8.A shows, the situation where the incomplete set of rules could be existed during the knowledge base is constructed. For instance; the constructed knowledge base contains the rule represented as follows;

| | |
|---|---|
| If the patient has viral hepatitis symptoms, | While; |
| The patient is with jaundice, | If he patient has viral hepatitis symptoms, |
| The HBsAg result is Reactive AND/OR | The patient with jaundice, |
| The HbsAg result Non-Reactive, | The HBsAg result Non-Reactive, |
| The IgM Anti HBc result is Reactive, Then | The IgM Anti HBc result is Reactive; |
| HBV result is Positive. | Then HBV result is not known. |

However, if such situations were occurred frequently the domain experts should fill the knowledge gap of the system through teaching or providing appropriate condition and action which also known as rule dynamically at runtime. The answer provided or validated by the expert is clearly shown in the figure 1.8.B above. Accordingly; when the knowledge gap of the knowledge base is filled by the domain expert, the knowledge base is validated and the provided answer is also saved to cope up with such situation for the future. As a result, to fill the knowledge gap of the system the domain expert also need to add the new condition "the result of HIV is positive" and conclusion "HBV result is positive" in to system knowledge base. Therefore; the rule is updated as:

*If the patient has viral hepatitis symptoms, the patient is with jaundice, the HBsAg result is Non-Reactive, the IgM Anti HBc result is Reactive, the result of HIV is positive; Then the HBV result is positive.*

## 9. IMPLEMENTATION

As we have seen the architecture in figure 9.1 below, the implemented adaptive learning expert system for the diagnosis and management of viral hepatitis consists several parts such as; the knowledge-base, inference engine, adaptive learner, user interface, and explanation facility.

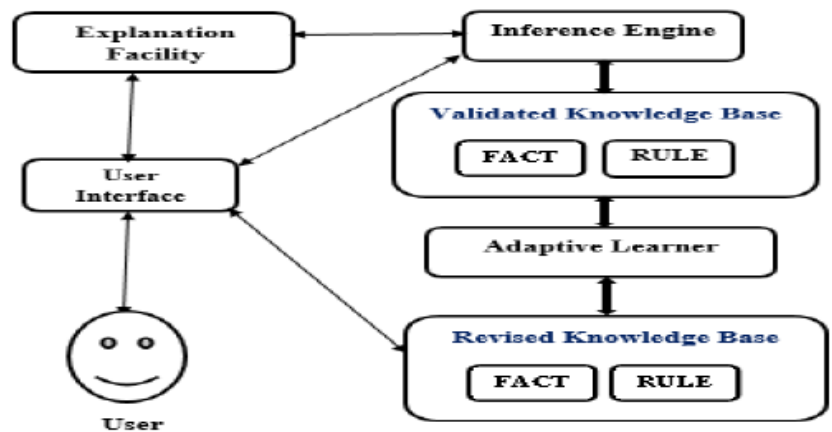

Figure 9.1 : Prototype Adaptive Learning Expert System Architecture

International Journal of Artificial Intelligence and Applications (IJAIA), Vol.10, No.2, March 2019## 10. CONCLUSION AND RECOMMENDATIONS

For the purpose of developing this adaptive learning expert system, tacit knowledge is acquired from domain experts by using structured and semi-structured interviewing techniques and also explicit knowledge is gathered from medical documents through document analysis technique. Once the appropriate knowledge is gathered, it is modeled by using decision tree to simulate procedures involved in diagnosis and management of viral hepatitis. Finally, the knowledge is represented using a rule-based knowledge representation methods and also SWI-prolog editor tool is used for codification of the represented knowledge. Among, those five (5) different types of viral hepatitis the scope of this study is limited on viral hepatitis B and viral hepatitis C only.

The following recommendations are forwarded based on the discusion of the findings of the study:

➢ In order to evercome each and every problems of the viral hepatitis the scope of the study should be extended.

➢ The system needs to be integrated with other languages like C#, VB, or Java.Net to have a more attractive look. Therefore, further research to design a user friendly user interface with the capability to use local languages is needed.

➢ Acquiring knowledge about the diagnosis and management of viral hepatitis is another challenge during knowledge acquisition from the domain experts using interviewing method. As a result, it is important to use the data mining tools and technics to filter out the hidden domain knowledge about the disease as a feature work of line.